# Transferability of Zr-Zr interatomic potentials

Oliver Nicholls[1], Dillion Frost[1], Vidur Tuli[1], Jana Smutna[2], Mark R. Wenman[2], Patrick A. Burr[1]*

[1] School of Mechanical and Manufacturing Engineering, The University of New South Wales, High Street, Sydney, NSW, 2052, Australia
[2] Department of Materials and Centre for Nuclear Engineering, Imperial College London, Prince Consort Road, Kensington, SW7 2AZ, United Kingdom

## Abstract

Tens of Zr inter-atomic potentials (force fields) have been developed to enable atomic-scale simulations of Zr alloys. These can provide critical insight in the in-reactor behaviour of nuclear fuel cladding and structural components exposed, but the results are strongly sensitive to the choice of potential. We provide a comprehensive comparison of 13 popular Zr potentials, and assess their ability to reproduce key physical, mechanical, structural and thermodynamic properties of Zr. We assess the lattice parameters, thermal expansion, melting point, volume-energy response, allotropic phase stability, elastic properties, and point defect energies, and compare them to experimental and ab-initio values. No potential was found to outperform all others on all aspects, but for every metric considered here, at least one potential was found to provide reliable results. Older embedded-atom method (EAM) potentials tend to excel in 2-3 metrics each, but at the cost of poorer transferability. The two highest-performing potentials overall, with complementary strengths and weaknesses, were the 2021 angular-dependent potential of Smirnova and Starikov (Comp. Mater. Sci. 197, 110581) and the 2019 embedded-atom method potential of Wimmer et al (J. Nucl. Mater. 532, 152055). All potentials trained through machine learning algorithms proved to have lower overall accuracy, and less transferability, than simpler and computationally faster potentials available. Point defect structures and energies is where the greatest divergence and least accuracy is observed. We created maps that will help modellers select the most suitable potential for a specific application, and which may help identify areas of improvement in future potentials.

# 1 Introduction

Zirconium (Zr) alloys have widespread use in water-cooled nuclear reactor cores due to their low thermal neutron capture cross-section, corrosion resistance and reasonable mechanical properties [1, 2]. Atomic scale simulations can provide valuable insight into the behaviour of Zr alloys under the extreme environments found in reactor cores. Molecular mechanics and molecular dynamics (MD) fill an important gap in the multi-scale modelling approaches that are required when studying multiscale phenomena, such as radiation damage [3]. However, the quality and reliability of MD simulations is only as good as the potentials used to describe the inter-atomic interactions, and while many potentials exist for Zr and Zr alloys, to the best of our knowledge there is no study that attempts to quantifies the degree of transferability and accuracy of these potentials.

Since 1988, 57 interatomic potentials for Zr have been developed [4-60]. Of these, 19 were developed for metallic glasses or thin films [4-22], 13 are general potentials for transition metals or complex alloys [23-37], and 23 were specifically designed for Zr alloys used in nuclear reactors [38-60]. In this study we are concerned with the latter category. Figure 1 shows the timeline of Zr potentials specifically designed for nuclear applications that were published between 1990 and 2021. Potentials developed before 1990 were superseded by the ones produce by Ackland, Wooding and Bacon [52], Mendelev and Ackland [54] and Kim, Lee and Baskes [56].

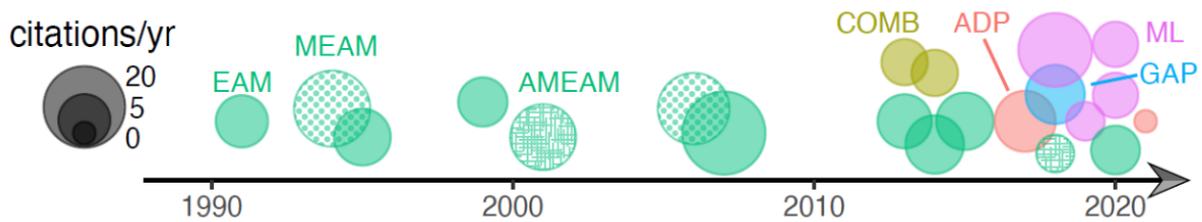

*Figure 1: History of Zr inter-atomic potential development between 1990 and 2021, coloured by potential type. Bubble size represented the average number of citations per year, as a metric of popularity.*

The number of relevant interatomic potentials developed per year has been increasing, particularly since 2012 (Figure 1). Many of the newer potentials build on previous potentials, improve performance, or extend their applicability, as highlighted in Table 1. This continual iteration and improvement of potentials, along with the significant diversity between predictions of even simple properties [61], highlights the need for consistent comparison between potentials.

This work studies the transferability of 13 Zr-Zr potentials [49-58] designed for simulations of Zr cladding and structural reactor components, detailed in Table 1 with the name with which they will be referred throughout this work. We first calculate the computation cost of each potential, and then we consider 15 metrics that can be directly compared with experimental or DFT data, to assess the transferability of the potential. These metrics are grouped 5 broad categories: dimer potential energies, thermodynamic phase stability, elastic properties, temperature-dependent structural parameters, and point defects. The potentials performance using these metrics is visualised in the Discussion section for ease of comparison. While there are other values comparison to be made between potentials (e.g. dislocation configurations, dipole tensors, gamma surfaces), the ones presented here were selected because they can be compared with exiting experimental or DFT literatures.

Table 1: Summary of potentials compared in this work. "SM" means the supplementary material for that paper.

| Family | Authors | Year | Name | Design criteria | Builds on | Sourced from |
|---|---|---|---|---|---|---|
| EAM | Ackland, Wooding and Bacon [52] | 1995 | EAM-AWB95 | α-Zr specific, Radiation damage | | NIST |
| EAM | Kim, Lee and Baskes [56] | 2006 | MEAM06 | Physical properties | Baskes and Johnson [44] | NIST |
| EAM | Mendelev and Ackland [54] | 2007 | EAM-MA07#2, | Phase transformation, Liquid structure | EAM-AWB95 | NIST |
| EAM | | | EAM-MA07#3 | Radiation damage | | |
| EAM | Lin, Wang, Peng, Li and Hui [53] | 2013 | EAM-LWP13 | Zr-Nb | | Tabulated |
| COMB | Noordhoek, Liang, Chiang, Sinnott and Phillpot [51] | 2014 | COMB14 | α-Zr specific, corrosion/oxidation Zr-O, Zr-H, Zr-O-H | Noordhoek, Liang, Lu, Shan, Sinnott and Phillpot [39] | SM |
| ADP | Smirnova and Starikov [49] | 2017 | ADP17 | Zr-Nb, $\omega$-Zr, Radiation Damage | | NIST |
| ML | Qian and Yang [57] | 2018 | GAP18#HCP, GAP18#BCC | Phonon Dispersion | | SM |
| ML | Zong, Luo, Ding, Lookman and Ackland [58] | 2019 | ML19 | α-Zr to $\omega$-Zr transformation | | From authors |
| EAM | Wimmer, Christensen, Wolf, Howland, Kammenzind and Smith [55] | 2019 | EAM-BMD19#1, EAM-BMD19#2 | Zr-H, thermal expansion, Radiation Damage | EAM-MA07 | From authors |
| ADP | Starikov and Smirnova [50] | 2021 | ADP21 | Zr-Nb, relative phase stability, radiation damage | ADP17 | NIST |

## 2 Methodology

All simulations were performed using LAMMPS [62]. The ML19 potential was run on the December 2015 release for compatibility reasons. Potential files were taken from the NIST repository [63], supporting information of the paper or directly from the author, as outlined in Table 1 with the exception being EAM-LWP13, which was tabulated using the *atsim.potential* python code of M.J.D Rushton [64]. The potential-supplied cut-off distances were used with an additional buffer layer of 2.0 Å. All MD runs were performed with a timestep of 2 fs. All static minimisations were performed with three consecutive conjugate gradient algorithms with a stopping tolerance of $10^{-25}$ relative energy change between two consecutive steps and maximum forces on atoms of $10^{-25}$ eV/Å.

For each potential we performed a consistent set of simulations to benchmark: (a) the computational cost of each potential; (b) the potential energy landscape of a Zr dimer; (c) the energy profile of a single atom displacement (d) the predicted stability of four Zr allotropes; (e) the elastic constants of α-Zr; (f) the temperature-dependent lattice parameters of α-Zr; (g) the melting point; and (h) the formation energy of intrinsic point defects. The computational cost was measured by running an MD simulation of a 10,976-atom HPC supercell for 1000 step with NVE ensemble. The dimer potential energy curve was calculated using an 80 x 80 x 80 Å box containing two atoms placed at a distance of 10 Å and progressively brought closer with a step size of 0.01 Å. The atomic displacement energy was profiled by moving one atom in a relaxed 180-atom HCP supercell in the $[11\bar{2}0]$, $[10\bar{1}0]$, $[0001]$ and $[2\bar{2}03]$ directions.

For the phase stability, we considered four allotropes of Zr: Hexagonal close packed (α), body centred cubic (β), face centred cubic (γ) and the hexagonal ω structure. For each phase, we calculate the ground state lattice parameters, the thermal expansion coefficient, and the change in energy with a change in volume of the unit cell, either by applying a hydrostatic strain (γ-Zr and β-Zr) or a hydrostatic pressure (α-Zr and ω-Zr), in increments of 0.02 % strain or 100 MPa, respectively. This approach was chosen so that the hexagonal crystal structure could relax to the lowest energy *c/a* ratio for each volume tested. Equivalent simulations were carried out with density functional theory (DFT) for comparison; these are described below. For the COMB14 potential, it was found that these distortions of the unit cell yield unstable simulations, and that the results were sensitive to supercell size used despite the periodic boundary conditions (details in supplementary materials). Thus, we report the results from a 3x3x3 supercell here, which appear to be the most consistent and reliable, and those from the unit cell and several other supercell sizes are reported in the supplementary materials.

The elastic constants were calculated by applying strains of $\pm 1 \times 10^{-6}$ in each of the six directions on a supercell of 864 atoms. The bulk ($B$) and shear ($G$) moduli of α-Zr were calculated using the Hill average ($B_H$ and $G_H$) [65, 66], of Voigt [67] and Reuss [68] approximations:

$$B_H = \frac{B_R + B_V}{2} \tag{1}$$

$$G_H = \frac{G_R + G_V}{2} \tag{2}$$

where the Voigt moduli $B_V$ and $G_V$ are:

$$B_V = \frac{2(c_{11} + c_{12}) + 4c_{13} + c_{33}}{9} \tag{3}$$

$$G_V = \frac{M + 12(c_{44} + c_{66})}{30} \tag{4}$$

with $M = c_{11} + 3(c_{11} - c_{12}) + 2c_{33} - 4c_{13}$; and the Reuss moduli $B_R$ and $G_R$ are:

$$B_R = \frac{c^2}{M} \tag{5}$$

$$G_R = \frac{5c^2 c_{44} c_{66}}{2[3B_V c_{44} c_{66} + c^2(c_{44} + c_{66})]} \tag{6}$$

where $c^2 = (c_{11} + c_{12})c_{33} - 2c_{13}^2$.

Changes in lattice parameter with temperature were simulated with an isothermal-isobaric ensemble (NPT) with 5-chain Nose-Hoover thermostat and barostat with damping times of 400 fs and 4 ps, respectively. The temperature was ramped from 20 K to 1750 K in increments of 10 K, except for the ML potentials which had a 20 K increment due to computational cost. At each temperature, the simulation was equilibrated for 4 ps followed by a 20 ps temperature hold during which data was collected. The coefficients of thermal expansions and the heat capacity were calculated using a central difference derivative with temperature of the lattice parameter and enthalpy, respectively. A 14-point average on either side of each point was taken to smooth data prior to differentiating:

$$\delta x_i = \frac{\sum_{i}^{i+14} x(i)}{15} - \frac{\sum_{i-14}^{i} x(i)}{15} \tag{7}$$

For the GAP18 and ML19 potentials we used a 7-point average due to the reduced number of temperature steps (thus the average covered the same temperature range).

The melting point, $T_m$, was calculated using the moving interface method (or coexisting phase method) [69, 70], with a 2400-atom supercell (5 x 5 x 24 α-Zr unit cells), equilibrated for 0.4 ns. If the solid/liquid ratio grew, the simulation temperature was deemed below $T_m$ and vice-versa if the liquid/solid ratio decreased. The simulations were repeated with three crystallographic planes parallel to the melting interface: $[11\bar{2}0]$, $[1\bar{1}00]$ and $[0001]$. The interface structure was created with the following procedure: after an initial NPT equilibrated of 20 ps, atoms in the lower half of the cell were frozen (no time integration) while those in the upper half were heated up to 2500 K in 5 ps, equilibrated for 10 ps, cooled back down to the target temperature in 5 ps, and equilibrated as a liquid for 20 ps. Throughout this process the barostat was applied to the z-direction only, to accommodate volume expansion, without applying in-plane strain to the solid phase.

Self-interstitial atoms (SIA) were simulated in supercells containing 360 lattice sites, at a constant volume that was scaled isotropically by $\frac{361}{360}$, to allow direct comparison with the equivalent DFT simulations of Vérité *et al.* [71]. Vacancy defects were also simulated with 360 lattice sites with one site unpopulated, the cell was allowed to relax to a minimum energy state. Defect relaxation volumes, $\Omega_{SIA}$, were calculated from the dipole tensor of the relaxed structure, and elastic constants of the material, with the aid of the Aneto code of Clouet [91].

One metric used in this work to compare the relative performance of potentials is the order of stability of Zr phases $\mathcal{O}(E_{allotrope}^f)$ or SIA defect structures $\mathcal{O}(E_{SIA})$. This is defined as the difference in rank of the phases or SIA structures, with respect to the that predicted by DFT, divided by two to avoid double counting

$$\mathcal{O}(E) = \frac{1}{2}\sum_i \left|rank(E_i^{pot}) - rank(E_i^{DFT})\right| \qquad (8)$$

Where two SIAs have the same energy, they are given the same rank and the next highest is skipped. Where a phase or SIA structure cannot be modelled by a potential it is assigned the maximum rank. Figure 11 and the graphical abstracts were created using the spider plot MATLAB add-on by M. Yoo [72].

Where experimental or DFT data was not available (e.g. energy-volume curves, defect relaxation volumes), additional DFT simulations were carried out using the VASP code [73, 74] with the PBE exchange correlation functional [75], projector augmented wave (PAW) pseudo-potentials and a consistent plane wave cut-off energy of 350 eV. Unit cells were modelled with Γ-centred k-point mesh with a linear spacing of 60 Å$^{-1}$, under hydrostatic pressure ranging from 9 GPa tension to 12 GPa compression. Cells were relaxed until the difference in total free energy between two electronic steps and two ionic steps was less than 10$^{-8}$ eV and 10$^{-6}$ eV respectively. Partial occupancies were treated with a first order Methfessel-Paxton smearing function of width 0.01 eV. Point defects were simulated using the same methodology of Vérité *et al.* [71], using a 360-atom supercell and 3x3x3 a k-point grid. Defect energies were in close agreement with those published previously [71], but the crowdion "C" configuration was found to relax spontaneously towards the lower-energy distorted crowdion C' configuration.

# 3 Results

## 3.1 Computational cost

The relative computational cost for each potential is shown in Figure 2. All EAM and ADP potentials are significantly faster than the other potential forms. The spread within the EAMs correlates with the cut-off distance of the potential, with longer cut-off requiring greater computational cost. COMB14 is approximately an order of magnitude slower, and the two potentials based on machine-learning techniques (ML19 and GAP18) are another order of magnitude slower again. In practical terms, this means that a simulation of a dislocation core with 100,000 atoms for 10 ps would take 4 minutes for MA07, 2.7 hours for COMB14 and over 188 hours (7.84 days) for the more recent ML potentials when run on 24 cores of a 2.7 GHz Intel Xeon Gold 6150.

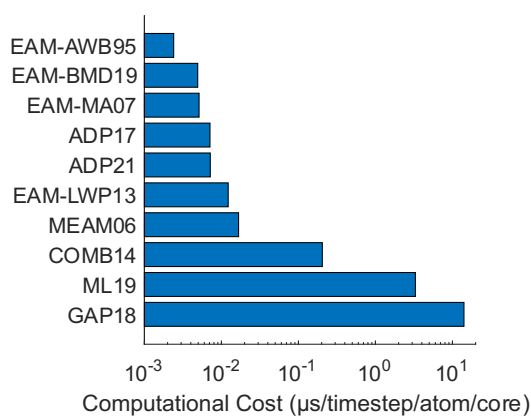

*Figure 2: Computational cost of each potential for a typical 100,000-atom MD simulation on a 2.4GHz CPU.*

## 3.2 Dimer potential energy

Figure 3 shows the energy profile of a Zr-Zr dimer molecule, as a function of bond length. The dimer potential energy profile shows a simple case where all *n*-body terms, embedding terms and angular dependencies are zero by construction, leaving only the pair-wise interaction. While this scenario is not directly relevant to solid state matter, and indeed none of the potentials studied here were explicitly fit to dimer potential energy, it is still illuminating to see the remarkable variation in inter-atomic interactions under this simple scenario. The GAP18 potentials exhibit very small inter-atomic energy at all distances, and with unphysically positive values for the α-Zr variant (i.e., repulsion at distances greater than 3.5 Å, but attraction at shorter distances). Additionally, as these potentials were not fitted to data at close inter-atomic separation, the energy of the dimer collapses to zero below 1.5 Å. This leads to highly unphysical results when two atoms are in close proximity to one-another, as is expected during radiation damage collision cascade event, thermal shocks, and high strain-rate deformations. COMB14 and MEAM06 exhibit a sharp collapse of the dimer energy at distances greater than 3.8 Å and 5 Å, respectively, indicative of the short-range nature of the pairwise component of these potentials. ADP potentials display a complex energy landscape, with several inflections and a region of repulsion around 4.28 – 6 Å. It is worth noting that the behaviour of a Zr-Zr dimer bares no direct consequences on the potential's performance on modelling solid phase Zr.

Unlike all other potentials considered here, the GAP formalism contains a chemical potential-like energy term that is intrinsic to the constituent atoms (i.e., an isolated GAP Zr atom has non-zero energy), which is –8.410 eV for the β-Zr variant and –8.439 eV for the α-Zr variant. To allow direct comparison with other potentials, the chemical potential was subtracted in the results presented in Figure 3.

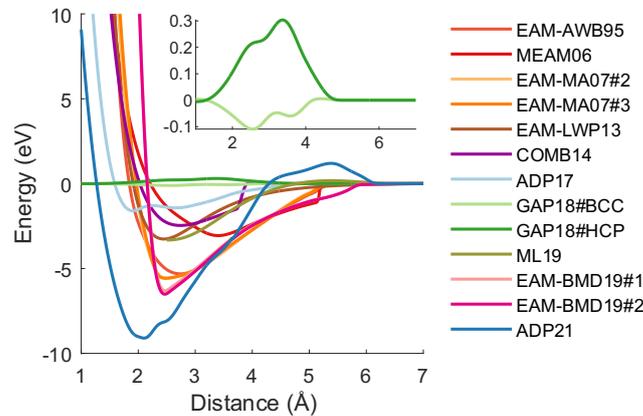

*Figure 3: Energy of Zr-Zr dimer as a function of interatomic distance, with respect to the energy at infinite separation. Inset: close-up of GAP18 potentials.*

### 3.3 Atomic displacement profile in HCP-Zr

The energy profile of a single atom being displaced within the lattice is shown in Figure 4 below. The two top panels show displacements towards the nearest neighbours: out-of-plane [2$\bar{2}$03] and in-plane [11$\bar{2}$0]. Most potentials perform indistinguishably between these two directions, suggesting no displacement anisotropy. Anisotropy of atomic displacements is expected in the HCP-Zr crystal, thus this is a potential area of further improvement for all potentials, which may have profound consequences on simulations of collision cascades and annihilation and clustering of radiation-induced defects. The other two directions considered here are towards the displacement to the 2$^{nd}$ and 3$^{rd}$ nearest neighbour, which require the moving atom to squeeze between (frozen) nearest neighbour atoms. This forma a characteristic hump at approximately half distance.

In the two primary directions, atomic displacements appear as smoothly decaying function for all potentials, without the oscillation observed at 0.4 fractional bond length in DFT. It is not clear whether this short-range feature in the DFT data is an artifact of the pseudopotential used in the calculations or it is a real feature. Further calculations using full-potentials and/or local basis sets would shed light on this, but that is beyond the scope of this work, since this feature is not of great interest even under high-energy collision cascade events: it is only observed at short range and when the neighbouring atoms are frozen in place, which is a highly artificial condition. Thus, when considering the relative performance of the potentials, we only compare the relative difference against DFT ion the fractional bond length range of 0.4–1, i.e. past the short-range DFT feature.

A simple EAM potential form is sufficient to capture the displacement energies accurately. The GAP18 potentials exhibit unphysically flat displacement profiles. The ML19 potential has discontinuities at 0.5 and 0.82 fractional bond length, which are also unphysical. Of the remaining potentials, the EAM-BMD potentials have by far the steepest displacement energy profiles of all potentials (overestimation of displacement energies), while the ADP potentials underestimate the displacement profile for fractional bond length < 0.6 (which is admittedly a fairly short-range interaction). It is interesting to note that the atomic displacement profiles are weak points for two of most versatile potentials in this study: the BMD19#2 and ADP21.

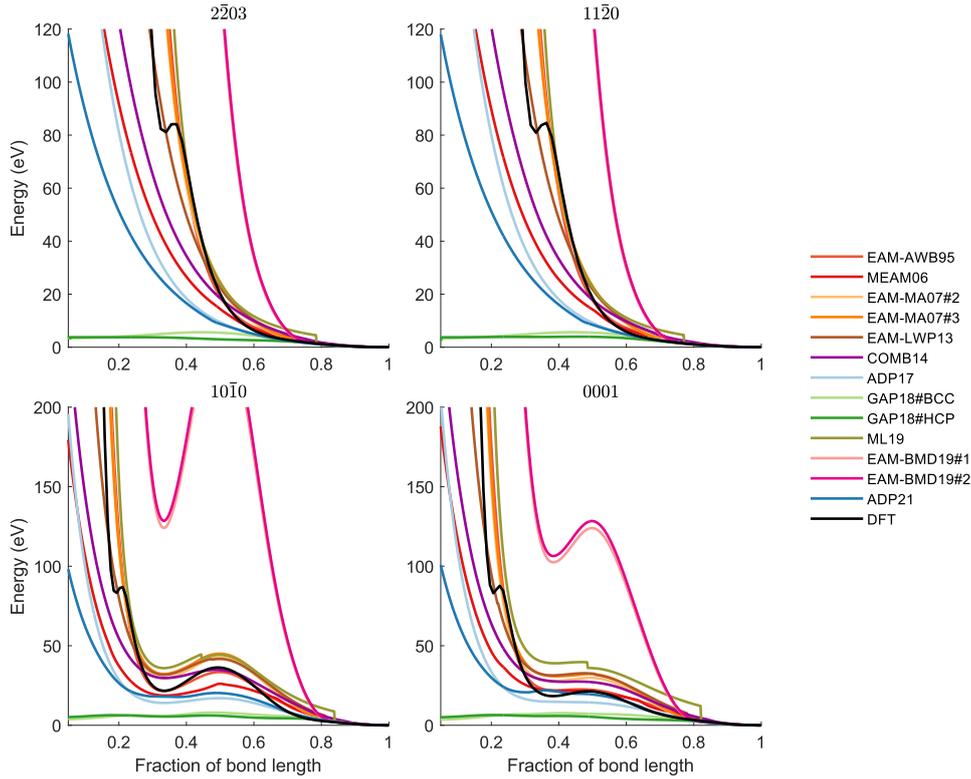

*Figure 4: Energy vs bond length for a single atom being dragged in a frozen α-Zr lattice.*

### 3.4 Phase stability of Zr allotropes

Next, we consider the thermodynamic stability of Zr allotropes (α-Zr, β-Zr, γ-Zr and ω-Zr). Here, it is important to highlight a long-standing discrepancy between *ab-initio* simulations and experiments regarding the ground state structure. The hexagonal ω-Zr phase is experimentally observed only under hydrostatic loads of > 2.2 GPa [76, 77], in samples that have undergone severe plastic deformation [77-82], and during shock loading [58, 83]. Conversely, DFT simulations predict the ω-Zr structure to be the ground state of Zr at 0 K [47], even when accounting for vibrational enthalpy and zero-point energy [47], with α-Zr having slightly higher energy (2 meV/atom [47]). This would imply, according to the DFT calculations, that ω-Zr is more stable, but its formation is kinetically hindered at low temperatures, and that α-Zr is thermally stabilised at some temperature below room temperature. However, experiments and DFT agree that β-Zr is only stable at high temperatures, forming above 1147 K at standard pressure [84], and at lower temperatures under pressure [76].

Figure 5 shows the difference in the ground state energy of Zr allotropes with respect to the energy of α-Zr for each potential. The absolute values for cohesive energies are presented in the Supplementary Material Figure S1. α-Zr is predicted to be the ground state structure for all potentials except for ADP17 and GAP19#HCP, which predict the β-Zr energy to be lower, and ADP21 which is the only potential to predict ω-Zr to have the lowest energy in agreement with DFT. It is worth noting that the stability of ω-Zr was part of the design considerations of the ADP21 potential [50]. Only EAM-AWB95, MEAM06, LWP13 and MBD19#2 do not predict β-Zr to be the highest energy structure. Importantly, GAP18#HCP and ADP17 predict the γ-Zr to have lower energy than α-Zr, which is not physical and may artificially induce excess stacking faults, twinning and phase transformations when modelling α-Zr. EAM-AWB95, EAM-MA07#2 and #3, COMB14, GAP18#BCC, ML19, EAM-BMD19#1 and #2, and ADP21 all retain the qualitative order of stability of $E_{HCP} < E_{FCC} < E_{BCC}$, although none accurately capture the relative energy difference of both $E_{BCC-HCP}$ and $E_{FCC-HCP}$, relative to DFT.

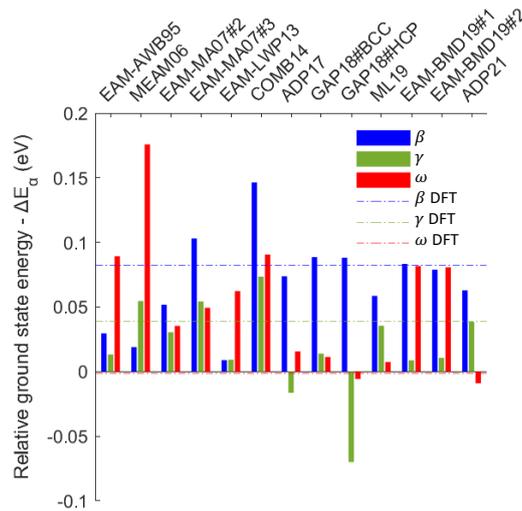

Figure 5: Ground state energy per atom relative to α-Zr for each potential considered in this study, compared to our DFT calculations (dashed lines).

### 3.5  Structure-energy-pressure relationship

Figure 6 shows the energy-volume relationship for each allotrope of Zr under applied hydrostatic pressure (hexagonal phases) or hydrostatic strain (cubic phases). The behaviour of each potential is compared with the DFT calculations of this work (hollow squares), aligned to have common energy minima, and each potential is offset vertically by 0.04 eV for ease of viewing. Where datapoints are missing, it implies that the potential relaxed into another structure during energy minimisation. An alternative format of the same data is provided in Figure S2, where the results are grouped by potential, which shows more clearly the hydrostatic pressure at which the phase transformations are predicted to occur.

In terms of ground state volume, most potentials accurately predict the unstrained α-Zr volume, likely because the potentials are fit to the ground state α-Zr lattice parameters, with the largest discrepancy seen in EAM-AWB95. The ground state volume of β-Zr is also accurately predicted by most potentials, with the largest discrepancy observed with COMB14. All potentials overestimate the volume for minimum energy of γ-Zr compared to DFT, with GAP18#HCP displaying an extreme case of this trend (molar volume 25 $Å^3$/atom). All EAM potentials predict a larger volume for minimum energy of ω-Zr compared to DFT (by a small margin in the case of BMD19#2), while GAP18#BCC strongly underestimates it. ADP potentials, GAP18#HCP and ML19 show good agreement with ω-Zr DFT data, likely because ω-Zr was part of the fitting dataset.

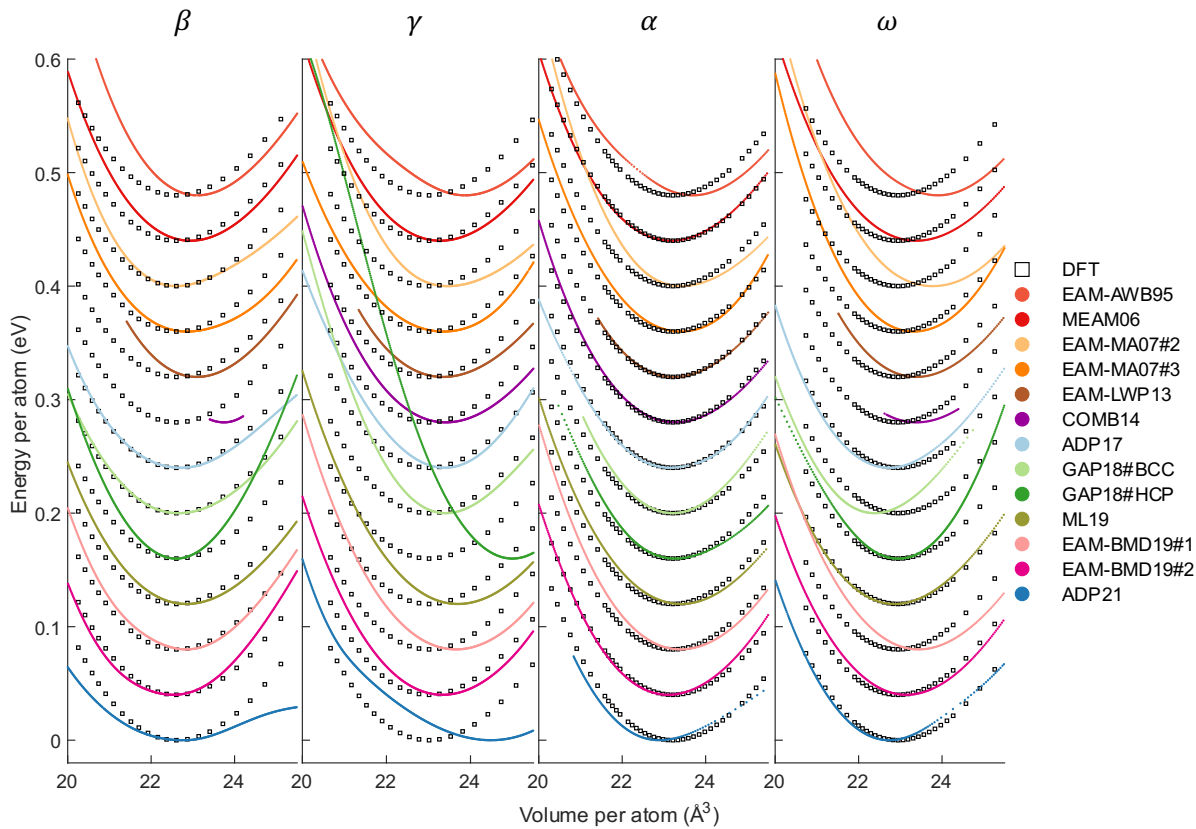

Figure 6: Energy-volume relationship of the four Zr allotropes (β-Zr, γ-Zr, α-Zr and ω-Zr), with a vertical offset of 0.04 eV between potentials. Square symbols represent the DFT value from this work, aligned so that the ground state of DFT and potential have the same energy.

Looking at the behaviour under compression and tension, MEAM06 has the most accurate prediction of the energy-volume relationship for α-Zr with ADP17, EAM-BMD19#2 and COMB14 also predicting closely. Conversely, the widely used EAM-MA07#2 exhibits remarkable discrepancy with DFT for the α-Zr structure under both compression and tension (remedied in EAM-MA07#3). GAP18#BCC most accurately predicts the energy-volume relationship of β-Zr, unsurprising given it is the only potential specifically designed for β-Zr. ML19 most accurately predicts the energy-volume relationship of ω-Zr (also unsurprising, as that was the goal of the potential) and EAM-BMD19#2 most accurately predicts the energy-volume relationship of γ-Zr. Where high stress or strain simulations are being conducted for non- α-Zr structures careful choice of potential is recommended given the high variability in performance.

The *c/a* ratio was allowed to change for ω-Zr and α-Zr during energy minimisation. DFT calculations inform that the α-Zr *c/a* ratio is expected to increase monotonically with increasing compression while the ω-Zr c/*a* ratio is expected to remain constant with increasing pressure [85]. Conversely, several potentials displayed non-monotonic, and in some cases discontinuous, behaviour for α-Zr and none maintained a constant ratio for ω-Zr. Further details are provided in the supplementary material Figure S3.

Under compressive and tensile strain, some of the potentials displayed unexpected or unphysical behaviour. Specifically, for EAM-AWB95 the ω-Zr phase transform to β-Zr at volume of 20.6 $Å^3$/atom, after displaying a very significant rise in energy; a rapid change in α-Zr *c/a* ratio (from 1.66 to 1.60) between 22.7 and 21.7 $Å^3$/atom is also observed for this potential. EAM-LWP13 has an approximate lower volume limit of 21.4 $Å^3$/atom, below which the potential fails to simulate all phases. Caution should be taken when using this potential in simulations with compression. GAP18#BCC yields a discontinuous step change in *c/a* ratio for the α-Zr structure at 21 $Å^3$/atom, from 1.60 to 1.45. This unphysical behaviour is perhaps irrelevant as this potential was designed for the β-Zr structure only. The α-Zr phase modelled by ADP21 has a step change in *c/a* ratio at 21 $Å^3$/atom from 1.59 to 1.54, and the β-Zr phase is predicted to have an inflection in the energy-volume relationship at high tension. Both EAM-AWB95 and ADP21 predict a change in γ-Zr behaviour at high compression. As mentioned in the methodology, COMB14's performance is strongly sensitive to super cell size. For the chosen super cell size (3x3x3), β-Zr and ω-Zr phases are only stable at near 0 stress, between 23.4 and 24.2 Å$^3$/atom and between 22.6 and 24.4 Å$^3$/atom, respectively. More information on the phase transformations of COMB14 is provided in Figure S4 of the supplementary material.

*3.6  α-Zr elastic constants*

The ability of each potential to describe the elastic properties of α-Zr was investigated by calculating the single crystal elastic constants, $c_{ij}$, and the Hill [65, 66] average of the Reuss and Voight bulk and shear moduli. These are shown in Table 2, where the colour of each cell represents the relative deviation from experimental value (also reported in brackets), together with the root mean square error of all $c_{ij}$ for each potential (colour-coded on a separate scale). A graphical visualisation is also provided in Figure 7. The EAM potentials generally appear to demonstrate the best performance in replicating the experimental elastic constants and moduli of α-Zr. ADP17 on the other hand displays poor performance, greatly improved in the more recent iteration, ADP21. Most potentials are within 10 % of the experimental value for bulk modulus except for the BCC-optimised variant of GAP18, which unsurprisingly provides poor representation of α-Zr elastic properties. Several potentials show significant deviation in elastic constants, but the resulting bulk modulus remains in close agreement with the experimental value (<5 % difference) which suggests the variation in elastic constants cancel out, particularly in cases where the potential parameters were fitted to the bulk modulus. $c_{66}$ is generally poorly represented, and this results in poor shear modulus, G, which is heavily based on $c_{66}$ [66].

Table 2: Elastic constants ($c_{ij}$), and bulk (B) and shear (G) moduli in GPa coloured by relative difference from experimental (indicated in brackets). Last column is the root mean squared error (RMSE) of the elastic constants with respect to experimental values, coloured by magnitude. * Indicates a quantity that was fitted to when the potential was developed. ** Indicate potentials where the fitted quantities are not stated.

| | $c_{11}$ | $c_{12}$ | $c_{13}$ | $c_{33}$ | $c_{44}$ | $c_{66}$ | B | G | RMSE % ($c_{ij}$) |
|---|---|---|---|---|---|---|---|---|---|
| Experimental | 144.0 [86] | 74.0 [86] | 67.0 [86] | 166.0 [86] | 33.0 [86] | 35.0 [86] | 95.9 [87] | 35.8 [87] | - |
| EAM-AWB95 | 151.3* (5.1%) | 85.2* (15.2%) | 69.8* (4.2%) | 174.6* (5.2%) | 35.5* (7.4%) | 33.0* (-5.6%) | 103 (7.4%) | 37.1 (3.7%) | 6.74 |
| MEAM06 | 151.5 (5.2%) | 71.8* (-3%) | 66* (-1.4%) | 160.6* (-3.2%) | 34.1* (3.4%) | 39.8* (13.8%) | 96.8* (1%) | 38.7 (8%) | 4.38 |
| EAM-MA07#2 | 148.3* (3%) | 81.6* (10.3%) | 63.3* (-5.5%) | 180.1* (8.5%) | 48.2* (46.1%) | 33.4 (-4.7%) | 99.2 (3.4%) | 42.9 (19.8%) | 9.33 |
| EAM-MA07#3 | 141.5* (-1.7%) | 74.2* (0.3%) | 74.0* (10.5%) | 167.7* (1%) | 43.9* (33.1%) | 33.6 (-3.9%) | 99.2 (3.4%) | 39.2 (9.5%) | 5.48 |
| EAM-LWP13 | 121.4* (-15.7%) | 86.7* (17.2%) | 77.9* (16.3%) | 178.4* (7.5%) | 27.7* (-16%) | 17.4 (-50.4%) | 99.7* (3.9%) | 25.1 (-30%) | 14.62 |
| COMB14** | 142.7 (-0.9%) | 84.3 (13.9%) | 58.9 (-12%) | 173.6 (4.6%) | 31.1 (-5.9%) | 29.2 (-16.6%) | 95.9 (0%) | 34.3 (-4.2%) | 6.68 |
| ADP17** | 157.6 (9.4%) | 136.1 (84%) | 66.8 (-0.3%) | 118.8 (-28.4%) | 8.5 (-74.3%) | 10.7 (-69.4%) | 103.2 (7.6%) | 13.8 (-61.4%) | 35.27 |
| GAP18#BCC | 197.5 (37.2%) | 86.8 (17.3%) | 93.5 (39.6%) | 190.5 (14.8%) | 26.7 (-19.2%) | 51.5 (47.3%) | 125.9 (31.3%) | 39.5 (10.3%) | 27.83 |
| GAP18#HCP | 157.7* (9.5%) | 64.3* (-13.1%) | 63.3* (-5.5%) | 191.2* (15.2%) | 15.5* (-53%) | 47.1 (34.7%) | 98.6 (2.9%) | 31.7 (-11.3%) | 15.21 |
| ML19** | 118.5 (-17.7%) | 90.1 (21.7%) | 77.7 (15.9%) | 136.9 (-17.5%) | 24.5 (-25.6%) | 14.2 (-59.4%) | 96.1 (0.2%) | 20.3 (-43.2%) | 9.23 |
| EAM-BMD19#1** | 142.8 (-0.8%) | 85.1 (15%) | 72.2 (7.8%) | 151.3 (-8.8%) | 30.4 (-8%) | 28.9 (-17.5%) | 99.6 (3.8%) | 31.5 (-12%) | 8.29 |
| EAM-BMD19#2 | 146 (1.4%) | 85.3 (15.2%) | 77.3 (15.3%) | 172.0 (3.6%) | 34.7 (5.1%) | 30.4 (-13.3%) | 104.7 (9.2%) | 34.6 (-3.5%) | 9.11 |
| ADP21** | 124.7 (-13.4%) | 97.8 (32.2%) | 75.1 (12.1%) | 150.8 (-9.2%) | 36.5 (10.5%) | 13.4 (-61.6%) | 99.6 (3.8%) | 24.6 (-31.4%) | 16.90 |

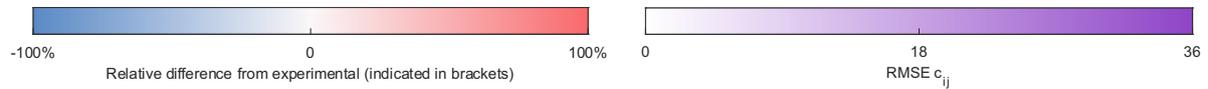

-100% — 0 — 100% Relative difference from experimental (indicated in brackets)

0 — 18 — 36 RMSE $c_{ij}$

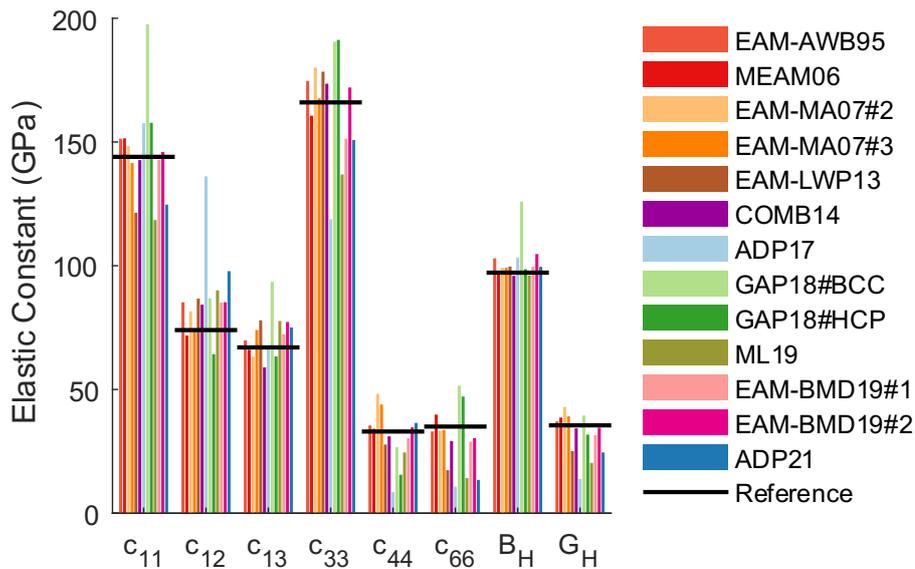

Figure 7: Elastic properties graphically visualised. Elastic constants reference: [86]. Bulk and shear modulus reference: [87].

## 3.7 Thermal expansion

The temperature dependence of the lattice parameters, volume, and *c/a* ratio of α-Zr was analysed by performing MD simulations with increasing temperature from 10 K until melting or spontaneous phase transformation. Figure 8 shows the evolution of the lattice parameters with temperature, and the resulting anisotropic coefficients of thermal expansion. The range of values obtained for the different potentials, and the widespread disagreement with experimental data (black line) is remarkable. ADP21 was the only potential to spontaneously transform to β-Zr (at 1470 K), while all other potentials melted, or in the case of GAP18#HCP, sublimated into a gas. Each point in Figure 8 represents the average of 10,000 steps (20 ps), and while the standard deviation at each point is not shown on the graph for clarity, this is reported in Figure S6 of the supplementary materials, as evidence that the simulation was sufficiently equilibrated after 20 ps. The coefficients of thermal expansion ($\alpha_a$, $\alpha_c$ and $\alpha_V$), being the derivative of the lattice parameters with respect to temperature, are sensitive to small fluctuations in the lattice parameters, which is apparent in the oscillations of selected potentials in Figure 8 (ADP21 seems particularly susceptible). Nevertheless, the current statistical sampling is adequate for the purpose of potential comparison.

Most potentials accurately predict the *a* and *c* lattice parameters to within 0.7 % of the experimental value, and slightly overestimate *c* (up to 1.5 %) over the temperatures for which α-Zr is stable. As a result, the volume and the *c/a* ratio are also generally overestimated, with the exception of MEAM06 and ADP21. The ADP21 potential stand out for overestimating the thermal expansion of α-Zr, but correctly capture the room temperature *a* (underestimating/overestimating at lower/higher temperatures). Conversely, GAP18#BCC, MEAM06 and EAM-LWP13 correctly exhibit constant $\alpha_a$. Notably, both of EAM-MA07, EAM-AWB95 and ADP21 exhibit an unphysical negative thermal expansion coefficient below 300 K in either *a* or *c* direction. This is also observed for volumetric expansion in the case of EAM-MA07#3 below 200 K.

In addition to the thermal expansion, we considered the enthalpy and heat capacity, and compared it to experimental value. All potentials perform poorly in this regard, with near-linear enthalpy change with temperature, and a slow increase in heat capacity with temperature, both of which deviate significantly from experimental trends. This is likely due to the intrinsic lack of an electronic system at this level of theory, irrespective of the potential form. For further details, refer to Figure S7 of the supplementary material.

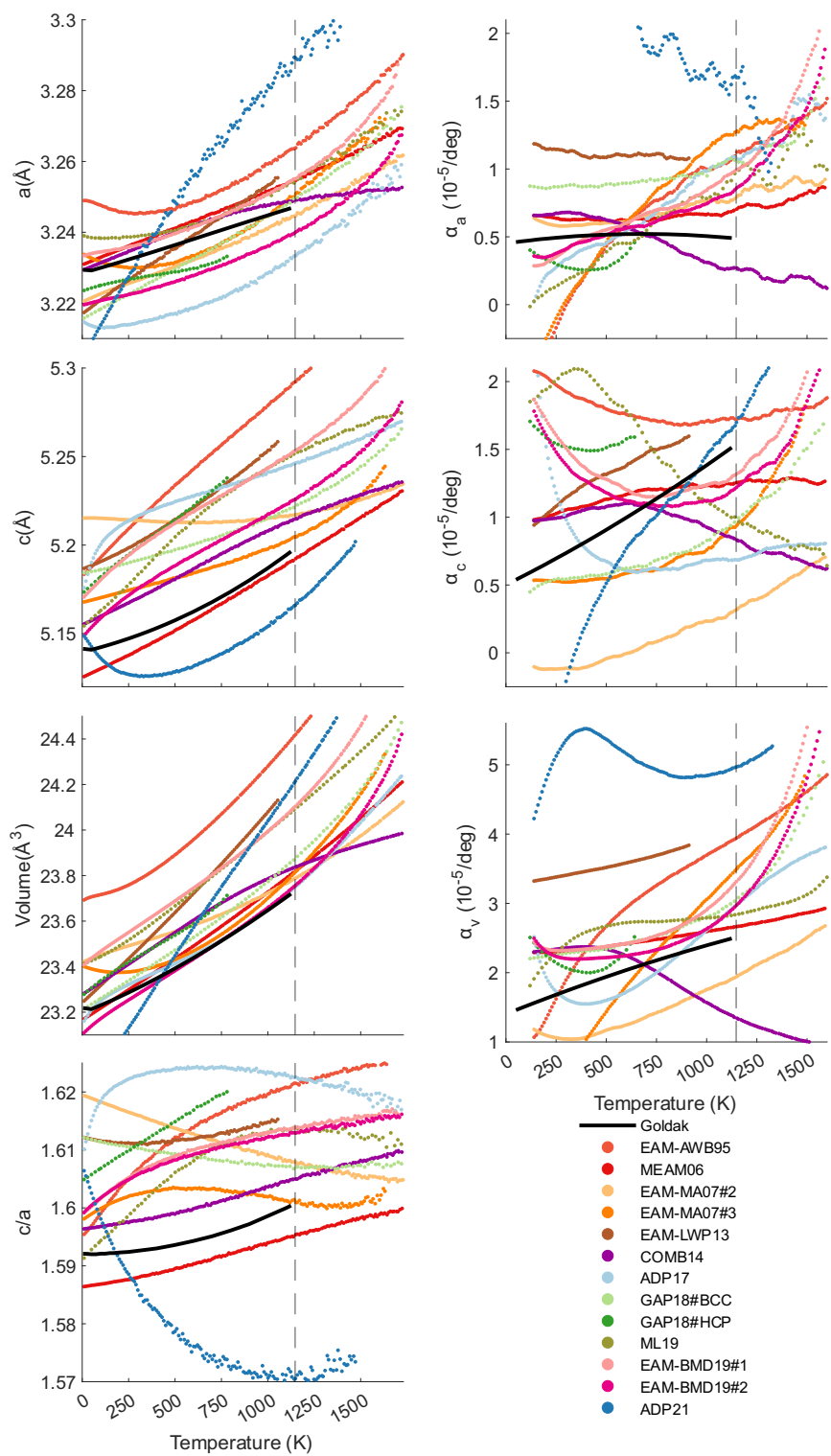

*Figure 8: Change in lattice parameters (left), and thermal expansion coefficient (right) of α-Zr with temperature. The vertical dashed line indicates the experimental transition temperature from α-Zr to β-Zr at 1147 K. The experimental values are taken from Goldak et al. [88].*

## 3.8 Melting point

The melting temperature of Zr, $T_m$, was analysed using the moving interface method [69, 70], for the $(11\bar{2}0), (1\bar{1}00)$ and $(0001)$ surfaces of $\alpha$-Zr. Figure 9 shows the predicted $T_m$ for each potential and compares it to the experimental melting point (2130 K [89]) and the reduced melting point of $\alpha$-Zr (1900 K [90]). The reduced melting point is an analytical estimation of the experimentally indeterminable melting point of low-temperature phases, in this case $\alpha$-Zr where it naturally melts from the $\beta$ phase [90].

The EAM and ML potentials, which were designed for α-Zr primarily, melt lowest. The MA07#2 potential was the only potential fitted to melting temperature, and unsurprisingly performs well in this test. All other potentials underestimate the melting point, with the MEAM06 and ADP21 performing better than the rest. This is especially true when considering the theoretical *reduced* melting point of α-Zr. Notably, MA07#2, ADP17 and ADP21 exhibit a spontaneous transformation of the solid phase from α-Zr to β-Zr, which is a testament to their ability to capture relative phase stability. The effect of crystal orientation appears to have limited impact on the melting point, providing further confidence of the results. A notable exception is ADP17, however this is due to the solid phase transformation to $\beta$-Zr in the cells that were originally oriented with the α-Zr $[1\bar{1}00]$ and $[0001]$ direction normal to the interface. No potential overestimated the melting point.

In some cases, we were unable to measure the melting point. Specifically, the MA07#2 $[11\bar{2}0]$ direction displayed unphysical behaviour of melting at two separate temperatures. It melted once from α-Zr at 1960 K, resolidified as β-Zr at 2060 K, and then melted again at 2120 K. The EAM-LWP13 displayed anomalous temperature transients of > 7000 K, which we were unable to control through the barostat. The COMB14 potential ran without issues up to 1900 K, however, above this temperature the $[11\bar{2}0]$ and $[0001]$ simulation proved to be unstable. A similar behaviour is observed for the [1$\bar{1}$00] orientation above 2400 K. Melting simulations of the GAP potentials are computationally very demanding and as such were not considered in this study.

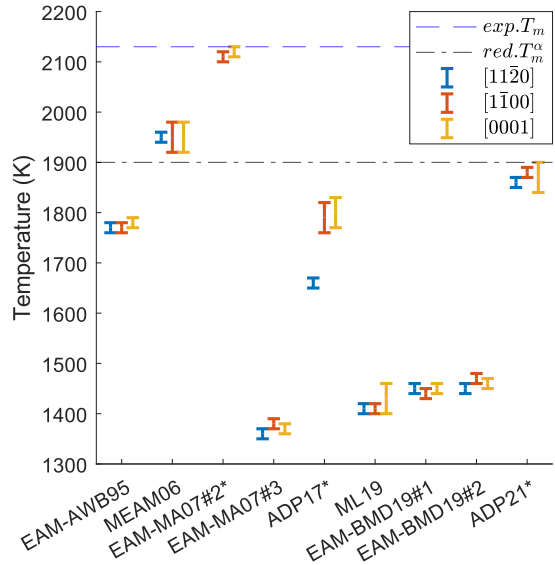

*Figure 9: α-Zr melting temperature calculated using the moving interface method with three crystal surface orientations. Potentials marked with * transformed to β phase before melting, except ADP17 in the [11$\bar{2}$0] direction which remained α until melting. Experimental melting point 2130 K ($exp.T_m$) from Liu et al. [89] and theoretical "reduced" melting point of α-Zr 1900 K ($red.T_m^\alpha$) from Ardell [90].*

## 3.9 Point defects

We investigate the potential's ability to model intrinsic point defects in α-Zr, specifically vacancies and self-interstitial atoms (SIA). It is well-established that point defect energy of Zr SIAs is strongly sensitive to supercell size [91-93]. Thus, to allow for a direct comparison with the reference DFT calculations, we simulated the defect using the same size supercell and same treatment of defect volume as the reference DFT study by of Vérité *et al.* [71]. Specifically they used a 360-atom supercell and performed constant-volume relaxations in a cell that was expanded isotopically by 361/360. As a result, the absolute values calculated here may differ from those calculated in other studies employing larger cells or a different treatment of defect volume, however these values are all internally consistent and therefore allow for direct comparison between potentials and with the DFT study.

Following the nomenclature of Vérité *et al.* [71] and Peng *et al.* [94], we consider the following SIA configurations: octahedral (O), basal octahedral (BO), tetrahedral (T), basal tetrahedral (BT), split dumbbell (S), basal split dumbbell (BS), prismatic (type-I) spit dumbbell (PS), crowdion (C), basal crowdion (BC), and the buckled version of the crowdions, C' and BC'. These are shown in Figure 10. These were used as initial geometries for our point defect calculations. It is important to note that the low-symmetry C' is sometimes also referred to an off-site octahedral or O'. Similarly, the distinction between BC and BO for the EAM-MA07#3 is subtle, and so is the difference between BS and BC for AWB95.

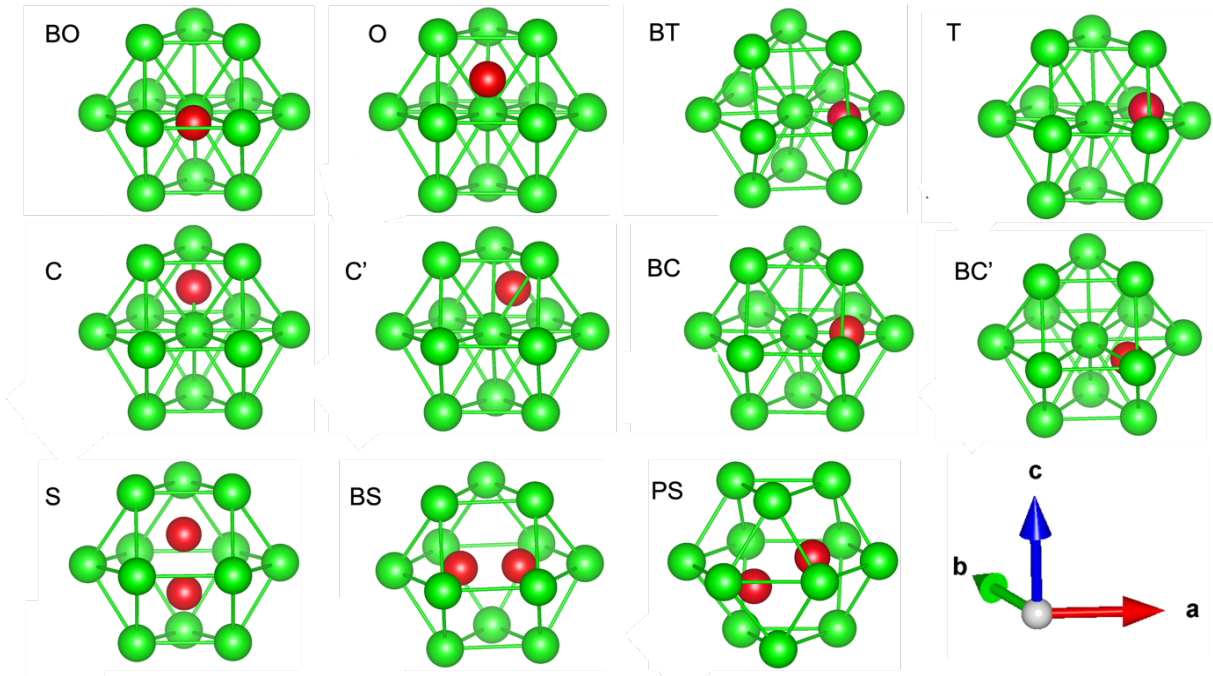

*Figure 10: pictorial representation of Zr SIA structures that have been proposed to be stable or metastable in the DFT literature.*

Table 3 shows the formation energy of isolated vacancies and SIAs in α-Zr, together with their combined energy to form a Frenkel pair, as this dictates the stored energy in the lattice following radiation damage events, and the RMSE of the SIA configurations compared to DFT, after removing one outlier with very high energy (S for ML19). Considering vacancies first, all potentials predict the vacancy formation energy with reasonable accuracy, with an average deviation of only -1.9 % and none were more than 21 % off, or 0.5 eV (ADP17). ADP21 has the closest agreement with DFT [71], closely followed by the BMD19 potentials, COMB14 and MEAM06.

*Table 3: Formation energy of SIAs ($E_{SIA}^f$), vacancy ($E_v^f$) and Frenkel pairs ($E_{FP}^f$) in eV, coloured by relative difference from DFT (indicated in brackets). † Indicates an unphysical collapse of the SIA onto another atom*

*such that they occupy the same coordinates. Root mean squared error (RMSE) for the SIAs coloured by magnitude.*

| | $E_{SIA}^f$ | | | | | | | | | | $E_v^f$ | $E_{FP}^f$ | RMSE$_{SIA}$ |
|---|---|---|---|---|---|---|---|---|---|---|---|---|---|
| | O | BO | S | BS | PS | C | C' | BC' | BC | T | BT | | | |
| DFT [71] | 2.84 | 2.73 | 2.93 | 2.82 | 2.95 | 3.15 | 2.95 | 2.75 | - | - | - | 2.0 [95] | 4.73 | |
| EAM-AWB95 | PS | 3.94 (44.4%) | BS | 3.91 (38.7%) | 3.88 (31.6%) | 3.96 (25.7%) | PS | BS | BS | BO | BS | 1.79 (-10.7%) | 5.67 (19.9%) | 0.72 |
| MEAM06 | 5.56 (95.8%) | 4.11 (50.6%) | 4.66 (59.2%) | 4.56 (61.7%) | 4.38 (48.3%) | 4.06 (28.9%) | 4.03 (36.5%) | 5.05 (83.6%) | BO | 4.62 | BC' | 2.1 (4.8%) | 6.12 (29.5%) | 1.76 |
| EAM-MA07#2 | C' | 2.89 (5.7%) | C' | BO | 3.15 (6.8%) | C' | 3.13 (6.2%) | BO | BO | PS | BO | 2.27 (13.5%) | 5.16 (9%) | 0.11 |
| EAM-MA07#3 | C' | 2.91 (6.7%) | C' | BC | 2.86 (-3%) | C' | 2.77 (-6%) | C' | 2.9 | BC | BC | 1.69 (-15.7%) | 4.46 (-5.7%) | 0.09 |
| EAM-LWP13 | PS | 3.35 (22.5%) | PS | BO | 3.34 (13.1%) | PS | PS | PS | BO | PS | PS | 1.68 (-16.1%) | 5.01 (6%) | 0.26 |
| COMB14 | 3.05 (7.3%) | O | PS | PS | 3.13 (6.1%) | 3.22 (2.3%) | 3.14 (6.6%) | O | 4.02 | 4.22 | 4.08 | 2.02 (0.8%) | 5.07 (7.1%) | 0.12 |
| ADP17 | C' | 3.09 (13.1%) | 3.49 (19.3%) | 3.2 (13.6%) | 3.18 (7.7%) | C' | 3.12 (5.8%) | C' | BO | S | BO | 1.58 (-21.1%) | 4.67 (-1.4%) | 0.29 |
| GAP18#BCC | 2.89 (1.9%) | 2.17 (-20.3%) | 2.43 (-17%) | NA | NA | S | NA | BO | BO | S | S | 2.15 (7.5%) | 4.32 (-8.6%) | 0.26 |
| GAP18#HCP | BO' | 1.32 (-51.5%) | 1.37 (-53.2%) | NA | NA | 1.79 (-43.3%) | S | BO | BO | S | S | 1.84 (-8.2%) | 3.16 (-33.2%) | 0.88 |
| ML19 | 2.8 (-1.4%) | 2.41 (-11.7%) | 29.23 (897.7%) | 2.55 (-9.7%) | 3.79 (28.5%) | 2.77 (-12.2%) | 2.85 (-3.3%) | BO | BO | 3.09 | BO | 2.3 (14.9%) | 4.71 (-0.4%) | 0.39 |
| EAM-BMD19#1 | 3.63 (27.9%) | 2.95 (8.2%) | PS | 3.16 (12.1%) | 3.22 (9.2%) | BO | PS | BO | PS | BO | BO | 2.03 (1.7%) | 4.99 (5.5%) | 0.33 |
| EAM-BMD19#2 | 3.5 (23.3%) | 2.89 (5.7%) | 3.54 (20.7%) | 3.11 (10.2%) | 3.13 (6.1%) | BO | PS | BO | BO | BO | BO | 2.08 (3.9%) | 4.97 (5%) | 0.34 |
| ADP21 | 2.28 (-19.6%) | 2.69 (-1.3%) | O | 2.76 (-2.1%) | 2.44 (-17.2%) | O | O | O | BO | BO | BO | 2 (0.2%) | 4.29 (-9.3%) | 0.27 |

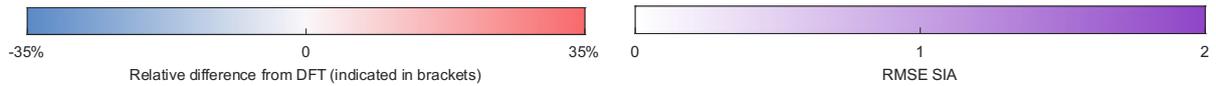

-35%   0   35% Relative difference from DFT (indicated in brackets)    0   1   2 RMSE SIA

When it comes to SIAs, all potentials perform poorly: most potentials exhibit only a few stable SIA configurations, which is in stark contrast with DFT simulations, which predict 8 distinct stable configurations. ML19, COMB14 and MEAM06, can predict at least 8 stable SIA configurations each, but these are not the same configurations that DFT predicts to be stable. ML16 is the only potential that can emulate the DFT structure, suggesting that these were part of the training set, however one configuration (S) yields unphysically high energy of 30 eV – this highlights the limited transferability of the potential despite being trained on the DFT point defect energies. Both GAP18 variants predict unphysical collapse of some SIAs onto the lattice site of another atom such that the two atoms occupy the same coordinates, and this is likely caused by the lack of inter-atomic repulsion below 1 Å.

As a metric of potential quality, we consider the root mean square difference in SIA energy between the potentials and DFT calculations. It is important to note that DFT energies of point defects often form part of the training set for the potentials, thus this metric isn't strictly a measure of transferability. Nevertheless, and despite this, we observe a surprisingly large disagreement between potentials and DFT. Overwhelmingly the formation energy of SIAs was higher than that predicted by DFT except for GAP18 potentials which predicted lower formation energies for nearly configurations. The average deviation from DFT reference value is 40 %. However, this metric can only be calculated for the SIA configurations that are predicted to be stable for both DFT and the potentials, and therefore it does not reflect how many of the structures are included in the comparison. For example, GAP18#BCC can only replicate two of the DFT SIA configurations, but for these two it predicts energies close to DFT and therefore this potential has the lowest RMSE value. This potential is closely followed by ADP 21. EAM-MA07#3, which was specifically designed to simulate point defects, predicts few stable SIAs, but with remarkably close agreement to DFT, if one considers that at the time the potential was developed most of the Zr SIA configurations had not been discovered [71, 94, 96]. If one considers the order of stability of the SIAs, rather than the absolute defect energy, then it appears that there is a general favouring of the BO site by most potentials, which is in agreement with DFT. O and PS configurations are also predicted to be low energy by most potentials, but these are not predicted to be low-energy sites by DFT: the O site is predicted to be the fourth most stable (out of 8), and PS is the least favourable.

An important quantity for calculations of radiation damage and radiation effects is the predicted amount of stored energy per Frenkel pair (a vacancy and SIA). Here we consider the energy required to form a dilute (non-interacting) Frenkel pair, composing the lowest energy SIA configuration. Under this definition, most potentials predict the Frenkel pair formation energy to within ±15 % of DFT, with notable exceptions of EAM-AWB95, MEAM06 (both > 6 eV) and GAP18#HCP (< 4 eV).

Finally, we consider the relaxation volume of point defects. This is a valuable metric because it is less often used during the training of potentials, and because it informs the elastic behaviour of point defects, as they annihilate or grow into larger defect such as dislocation loops. Table 4 shows the volume of stable point defects, colour-coded by the difference in volume compared to the DFT calculations carried out in the same supercell and with the same relaxation methodology. Notably, no potential is in close agreement with DFT, with typical errors ranging in the ±40%, and in two cases larger than 100%. The BMD19 potentials are the highest performing potentials in this category, but even these have two defect configurations with errors exceeding 20%. Most potentials tend to underestimate the relaxation volume of vacancies, and only the COMB14 comes close to the DFT-calculated value for vacancies. Some potentials underestimate the volume of all defects (AWB95, MEAM06, LWP13 and COMB14), while others have errors on both sides of the scale. Interestingly, the ADP17 potential exhibits closer agreement with DFT in this regard than its successor, the ADP21 potential. The ML19 potentials performs exceptionally well for all point defects except the split configuration "S", for which it predicts an unphysical reduction in volume. This highlights the key shortcoming of machine-learning potentials: they excel in simulating cases that are close to or part of the learning dataset, but can fail spectacularly and unexpectedly for cases that fall beyond the transferability regime, despite appearing relatively similar to the training dataset.

Table 4: Relaxation volume of SIAs ($\Omega_{SIA}$), vacancy ($\Omega_v$) in Å$^3$, coloured by relative difference from DFT (indicated in brackets). † Indicates an unphysical collapse of the SIA onto another atom such that they occupy the same coordinates. Root mean squared error (RMSE) is coloured by magnitude.

| | $\Omega_{SIA}$ | | | | | | | | | | $\Omega_v$ | RMSE$_\Omega$ |
|---|---|---|---|---|---|---|---|---|---|---|---|---|
| | O | BO | S | BS | PS | C | C' | BC' | BC | T | BT | | |
| DFT | 21.69 | 22.06 | 23.74 | 21.50 | 21.61 | | 21.83 | 22.24 | | | | -10.83 | |
| EAM-AWB95 | PS | 17.56 (-20.4%) | BS | 16.17 (-24.8%) | 15.53 (-28.1%) | 18.03 | PS | BS | BS | BO | BS | -6.32 (-41.6%) | 5.15 |
| MEAM06 | 12.99 (-40.1%) | 14.17 (-35.8%) | 15.72 (-33.8%) | 12.81 (-40.4%) | 14.36 (-33.6%) | 13.88 | 14.75 (-32.4%) | 14.54 (-34.6%) | BO | 14.26 | BC' | -7.79 (-28.1%) | 7.49 |
| EAM-MA07#2 | C' | 27.89 (26.4%) | C' | BO | 29.93 (38.5%) | C' | 29.83 (36.7%) | BO | BO | PS | BO | -1.1 (-89.8%) | 8.09 |
| EAM-MA07#3 | C' | 29.85 (35.3%) | C' | BC | 27.14 (25.6%) | C' | 25.84 (18.4%) | C' | 27.94 | BC | BC | -8.78 (-18.9%) | 5.28 |
| EAM-LWP13 | PS | 7.9 (-64.2%) | PS | BO | 9.7 (-55.1%) | PS | PS | PS | BO | PS | PS | -9.42 (-13.0%) | 10.71 |
| COMB14 | 16.07 (-25.9%) | O | PS | PS | 15.85 (-26.6%) | 16.14 | 15.2 (-30.4%) | O | 14.54 | 12.09 | 13.68 | -10.76 (-0.6%) | 5.21 |
| ADP17 | C' | 24.27 (10.0%) | 10.24 (-56.9%) | 23.16 (7.8%) | 18.8 (-13.0%) | C' | 19.95 (-8.6%) | C' | BO | S | BO | -12.51 (15.6%) | 5.83 |
| GAP18#BCC | 31 (42.9%) | 32.33 (46.5%) | 27.01 (13.8%) | † | † | S | † | BO | BO | S | S | -5.29 (-51.1%) | 7.64 |
| GAP18#HCP | BO' | 27.16 (23.1%) | 11.46 (-51.7%) | † | † | 20.03 | S | BO | BO | S | S | -3.76 (-65.3%) | 8.69 |
| ML19 | 20.86 (-3.9%) | 24.21 (9.7%) | -4.07 (-117%) | 21.87 (1.7%) | 18.96 (-12.3%) | 22.23 | 22.2 (1.7%) | BO | BO | 24.71 | BO | -12.48 (15.3%) | 10.61 |
| EAM-BMD19#1 | 26.44 (21.9%) | 21.84 (-1.0%) | PS | 19.43 (-9.6%) | 23.55 (9.0%) | BO | PS | BO | BO | PS | BO | -8.17 (-24.5%) | 2.75 |
| EAM-BMD19#2 | 26.52 (22.3%) | 21.77 (-1.3%) | 28.09 (18.3%) | 19.47 (-9.4%) | 23.44 (8.5%) | BO | PS | BO | BO | BO | BO | -7.29 (-32.7%) | 3.22 |
| ADP21 | 21.3 (-1.8%) | 34.67 (57.1%) | O | 27.26 (26.8%) | 27.35 (26.6%) | O | O | O | BO | BO | BO | 5.18 (-148%) | 9.81 |

-100% — 0 — 100%  Relative difference from DFT        0 — 6 — 12  RMSE (Å$^3$)

## 4 Discussion

To synthesise the performance of the potentials considered, a range of key quantitative indicators were chosen. Where possible, these were compared with experimental values, alternatively DFT simulations were used for benchmarking. For structural parameters we consider four metrics: relative difference with experimental values of room temperature lattice parameters (290 ± 10 K), root mean square error (RMSE) of the same calculated over the temperature range 4.2-1100 K, reported separately for $\Delta a_T$ and $\Delta c_T$ and the melting temperature difference from experimental 2130 K [89] ($\Delta T_m$). The performance in terms of mechanical response is embodied in the RMSE of the elastic constants ($\Delta c_{ij}$) with respect to experimental values (as reported in Table 2), and in the energy response to hydrostatic deformation of each Zr allotrope compared to DFT data ($\Delta E(V)_{allotrope}$). For radiation damage, we consider five metrics: The absolute percentage error from DFT in Frenkel pair formation energy ($\Delta E^f_{FP}$), the RMSE of the SIA formation energy (only considering those SIA configurations that can be reproduced in DFT)($E^f_{SIA}$), the order of the formation energy of the SIA configurations ($\mathcal{O}(E_{SIA})$) defined in eq. (8), the RMSE of the relaxation volume ($\Delta\Omega_{SIA}$), and the fractional RMSE of the energy profile of a single atom displacement averaged over four directions $\langle\Delta E_d(x)\rangle_\theta$. The $\Delta E_d(x)$ for each of the directions individually can be seen in Figure S8. The two EAM-BMD19 potentials have a nearly 4x larger RMSE compared to the next potential (3.81 and 3.95 respectively), to allow for differentiation between the other potentials EAM-BMD19's results have not been included in Figure 11. For thermodynamic stability we consider the RMSE of the formation energies for all allotropes against DFT values ($\Delta E^f_{allotrope}$) and the order of stability of the allotropes ($\mathcal{O}(E^f_{allotrope})$) is calculated using eq. (8).

Figure 11 shows the rankings of the potentials for each of the indicators described above. The potentials are grouped by type: the top row are EAMs split with the potentials published before/after 2010 on the left/right. The bottom left panel comprises ADP and COMB14 potentials and the bottom right are potentials based on machine learning approaches (ML and GAP).

Overall, older EAM were more specialised, ranking highly in some areas (e.g., 273 K lattice parameters) and poorly in others (e.g., $\Delta E(V)_\omega$ and $\mathcal{O}(E_{SIA})$), while more modern EAM potentials are more generalist potentials with more consistent performance throughout. Most potentials (especially the EAMs) perform well for elastic constants with the notable exceptions of ADP17 and GAP18#BCC. Surprisingly, the GAP18#HCP potential appears to perform worse than the GAP18#BCC overall, despite many of the indicators being specific to the α-Zr structure. This suggest that the GAP potential form may have significant space for improved if it were fitted to a greater range of relevant quantities. While the COMB14 potential appears to have significant areas of weaknesses, one must also consider that this potential is specifically designed to model charge transfer between metallic and ionic valences of Zr in the presence of O and/or H. Conversely, the test performed in this study are biased towards potentials explicitly designed for metallic state of Zr. ADP21 is by far the best performing potential for phase stability and has significant improvements in the thermal and elastic parameters over ADP17 but at the cost of point defect performance. ML19 stands out in predicting point defect structures and relative energies (order of stability of SIA configurations), with the closest agreement to DFT – as would be expected when including point defect energies in the training set.

ADP21 and EAM-BMD19#2 stand out as a particularly well-rounded potentials, with mostly complementary strengths. Areas where there is considerable space for improvement are thermal expansion (for both), displacement energies (for both), point defect stability (for both), phase stability and melting point (for BMD19#2), elastic constants and point defect volumes (for ADP21). ML19 and EAM-MA07#3 also show strong performance in most categories, however, have mediocre performance in one or two areas each. However, ML19 has shown to have unpredictable and unphysical behaviour when atomic configurations deviated from the training set, as is expected of machine learning potentials used beyond their design scope. It is important to bear in mind that point defects, elastic constants and phase stability are often part of the fitting strategy when generating the potentials, and as such the apparent good agreement with DFT or experimental values observed here should not be interpreted as a good measure of transferability. On the contrary, it is surprising how poor most potentials perform in predicting even basic quantities, such as thermal expansion and point defect geometries, when these were part of the training set of the potentials.

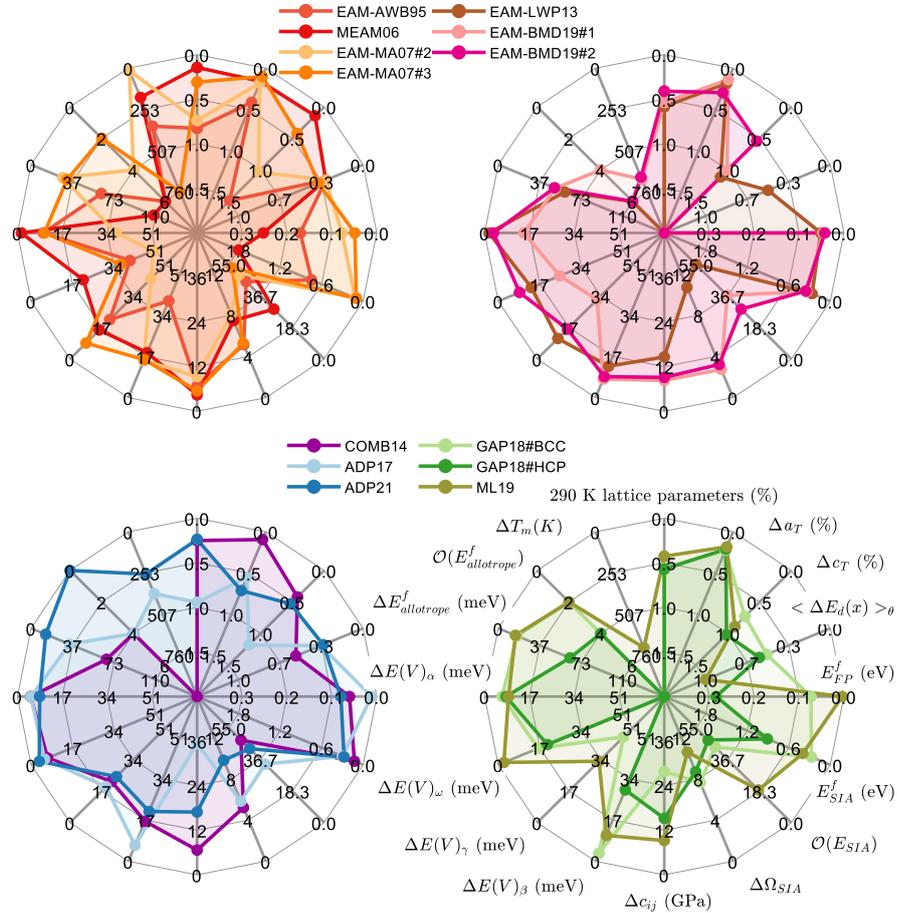

*Figure 11: Radar plot showing the RMSE performance of the potentials in key metrics. Top left: EAMs pre-2010. Top right: EAM post-2010. Bottom left: ADP and COMB. Bottom right: Machine learnt potentials. Potentials for which the melting point could not be determined are marked at the centre of their respective plots in the $\Delta T_m$ axis. GAP18#HCP $\Delta E(V)_{FCC}$ has an error of 184 meV, beyond the range of the scale.*

This study highlights that for every application considered here, there is at least one potential currently available that performs with reasonably close agreement to either experimental data (where available) or DFT data. However, no potential proved to be universally transferable, even within the relatively small set of performance metrics considered here. Thus, the choice of potential must be taken with great care, and with specific consideration to the set of properties required for the study, as there is substantial level of disagreement between different potentials. Even greater care must be taken when a study requires a potential with good description of a combination of properties, for example a study on radiation damage at temperature would require a good description of point defects, thermal expansion, and relative phase stability, and no one potential significantly outperformed all others when considering all three aspects. The study also highlights the areas with largest areas for improvement (e.g. point defects), which we hope will guide the development of future potentials for Zr and Zr alloys.

## 5 Conclusion

In this work we have compared the performance of 13 zirconium potentials and evaluated their capacity to accurately simulate a series of parameters relevant to nuclear applications. No individual potential showed outstanding performance on all accounts, but for every property at least one potential performed with adequate accuracy. However, the discrepancy between potentials, even for relatively fundamental quantities, is remarkable, thus great care must be taken to choose the appropriate potential for the specific set of properties that are relevant to each study.

All potentials showed less than 1 % error on the room temperature lattice parameters. The agreement remains admirable when considering thermal expansion in the range 140-1125 K, however, this is better for $a$ lattice parameter (<0.5 % RSME) than the $c$ lattice parameter (>1 % RSME). Interestingly, the older EAM potentials performed best on the elastic constants. Most potentials struggled to accurately predict the relative formation energy for the phases relative to α-Zr, with the 2021 ADP potential of Smirnova and Starikov performing the best in this metric. With the exception of machine-learning potentials, newer potentials predict the relative energy of Zr allotropes and their change in energy with change in volume more accurately, likely due to better fitting for ω-Zr and γ-Zr. By extension, when choosing a potential to simulate non- α-Zr structures under stress or strain careful consideration should be made due to the inconsistency across the phases. The 2007 EAM by Mendelev and Ackland #2 is the only potential to capture the experimental melting point of Zr, however ADP21 and the MEAM of Baskes also get close to the theoretical reduced melting temperature of α-Zr, while all other potentials underestimate the melting point by several hundreds of degrees. Most potentials predicted the formation energy of vacancies and of Frenkel pair with less than 16 % error, however, all potentials struggled to accurately predict the structure, formation energy and defect volumes of SIAs, with BMD19, ADP19/21 and ML19 having the least (but still substantial) deviation from DFT.

Overall, simpler potential forms such as EAM and ADP, which are orders of magnitude faster than recent machine-learning potentials, appear to outperform the more complex potentials in most metrics considered here. There is no clear potential that can be recommended for universal use as each potential has a few significant limitations. The two most all-rounded potentials are ADP21 and EAM-BMD19#2, and their strengths are complementary. A clear opportunity for further development of interatomic potentials for Zr lies in the ability to predict potin defects, and by extension locally deformed structures, such as those that may be do found at core of dislocation and near extended defects.

## Data availability

The tabulated potentials used, and raw data to reproduce these findings is available at [INSERT DOI TO DATA ARCHIVE]

## Acknowledgements

This research was supported by two Australian Government Research Training Program (RTP) Scholarships. PB and VT would like to thank Westinghouse Electric (Sweden) for providing financial support. This work was undertaken with the assistance of resources and services from the National Computational Infrastructure (NCI), which is supported by the Australian Government; the Multi-modal Australian ScienceS Imaging and Visualisation Environment (MASSIVE); the Pawsey Supercomputing Centre, which is supported by the Australian Government and the Government of Western Australia; and was enabled by Intersect Australia Limited. Learn more at www.intersect.org.au. The authors thank Dr. Rushton, for sharing the atsim.potential python code [64], and we extend special thanks to Moses Yoo for the timely and generous support with the MATLAB spider plot add-in [72] used to create Figure 11.